\documentclass[traditabstract,twocolumn]{aa}

\usepackage{epsfig,graphicx,natbib}
\usepackage{url}
\usepackage{longtable, amssymb}
\usepackage{ulem}

\begin{document}

\title{A sub-ppm upper limit on the cosmological variations of the fine structure constant $\alpha$}

\author{S.~Muller \inst{1}
\and A.~Beelen \inst{2}
\and M.~Gu\'elin \inst{3,4}
\and J.\,H.~Black \inst{1}
\and F.~Combes \inst{5}
\and H.\,L.~Bethlem \inst{6}
\and M.~G\'erin \inst{4}
\and C.~Henkel \inst{7}
\and K.\,M.~Menten \inst{7}\thanks{In memoriam}
\and M.T.~Murphy \inst{8}
\and W.~Ubachs \inst{6}
\and N.~Wozny \inst{9}
}

\institute{Department of Space, Earth and Environment, Chalmers University of Technology, Onsala Space Observatory, SE-43992 Onsala, Sweden
\and Aix-Marseille Univ, CNRS, CNES, LAM, Marseille, France
\and Institut de Radioastronomie Millim\'etrique, 300, rue de la piscine, 38406 St Martin d'H\`eres, France 
\and LRA/LUX, Observatoire de Paris, PSL, CNRS \& \'Ecole Normale Sup\'erieure, 75231 Paris, France
\and LUX, Observatoire de Paris, PSL, Coll\`ege de France, CNRS, Sorbonne University, 75014 Paris, France 
\and Department of Physics and Astronomy, VU University Amsterdam, De Boelelaan 1100, 1081 HZ Amsterdam, the Netherlands
\and Max-Planck-Institut f\"ur Radioastonomie, Auf dem H\"ugel 69, D-53121 Bonn, Germany
\and Centre for Astrophysics and Supercomputing, Swinburne University of Technology, Hawthorn, Victoria 3122, Australia
\and Institut d’Astrophysique de Paris, UMR 7095, CNRS, Sorbonne Université, 98 bis boulevard Arago, F-75014 Paris, France
}

\date {Received / Accepted}

\titlerunning{A sub-ppm upper limit on the cosmological variations of the fine structure constant $\alpha$}
\authorrunning{Muller et al. 2025}

\abstract{Absorption spectroscopy toward high-redshift quasars provides strong constraints on the putative variation of fundamental constants of physics on cosmological time scales. The submillimeter ground-state transitions of methylidyne (CH) and water (H$_2$O), both molecules widespread and coeval in the interstellar medium, provide a sensitive test for variations of $\alpha$, the fine structure constant, and $\mu$, the proton-to-electron mass ratio, taking advantage of the unmatched spectral resolution and frequency reliability of radio techniques. We used ALMA simultaneous observations of the two species to constrain any velocity offset between their absorption profiles toward the radio-bright lensed quasars PKS\,1830$-$211 ($z_{\mathrm abs}=0.88582$) and B\,0218$+$357 ($z_{\mathrm abs}=0.68466$). Our observational setup minimizes instrumental errors and known sources of systematics, such as time variability of the absorption profile and frequency-dependent morphology of the background quasar. The excellent correlation between CH and H$_2$O opacities, the large number of individual narrow velocity components, and the number of independent spectra obtained due to the intrinsic time variability of the absorption profiles ensure that even the chemical segregation bias is minimized. We obtained bulk velocity shifts $\delta v = -0.048 \pm 0.028$~km\,s$^{-1}$ and $-0.13 \pm 0.14$~km\,s$^{-1}$ (1$\sigma$ confidence level) between CH and H$_2$O in the direction of PKS\,1830$-$211(NE) and B\,0218$+$357(SW), respectively. These values convert into the $3\sigma$ upper limits $|\Delta \alpha / \alpha| < 0.55$~ppm and 1.5~ppm, respectively, taking into account the independent upper limits on $|\Delta \mu / \mu|$ previously obtained for these systems. These constraints on $|\Delta \alpha / \alpha|$, at look-back times of about half the present age of the Universe, are two to four times deeper than previous constraints on any other single high-$z$ system.
}


\maketitle

\section{Introduction} \label{sec:intro}

Whether new physics exists beyond the Standard Model of Particle Physics and General Relativity remains one of the most compelling questions in modern science. Einstein's Equivalence principle, which states that the result of any non-gravitational experiment is independent of when and where it is performed, has been confronted for a century to deep investigations, involving atomic clocks, spontaneous fission in the Oklo natural reactor, and, lately, astronomical observations. So far, only hints of a possible violation of the Equivalence principle have been reported, essentially based on measurements of the fundamental fine-structure constant $\alpha$ in remote quasars (e.g., the reviews by \citealt{mar17,uza25}), although they are now most likely explained by systematic errors (e.g., \citealt{rah13,whi15}).

The Equivalence Principle was challenged in the 1930's by \cite{dir37}, who postulated that dimensionless ratios of universal fundamental constants, such as the proton-to-electron mass ratio $\mu = m_{\rm p}/m_{\rm e} = 1836.152673426(32)$ and the electromagnetic fine-structure constant $\alpha = e^2/ 4\pi\epsilon_0\hbar c = 7.2973525643(11) \times 10^{-3} \sim 1/137$ (2022 CODATA recommended values \footnote{\url{https://physics.nist.gov/cuu/Constants/index.html}}, \citealt{moh25}), are not plain numbers, but have profound significance and may vary through space-time. The question gained special attention after the discovery, in 1998, that the expansion of the Universe is accelerating since at least 6~Gyr \citep{rie98}. This acceleration, which is expected to be sparked by an enigmatic {\it dark energy} \citep{fri08}, renewed interest in post-Einsteinian gravitational models involving extra dimension(s) to those of space-time. From there, a {\it quintessence} or {\it phantom} dynamical scalar field would couple with the charged matter and induce time and/or space variations of the electromagnetic interaction between charged particles, hence variations of $\alpha$. Accurate measurements of the $\alpha$ value in remote galaxies allow us to test this hypothesis (e.g., \citealt{sav56, bah67,dri98}). From a theoretical point of view, variations of $\mu$ and $\alpha$ are established to be correlated in models of Grand Unification, although the correlation coefficient is highly model dependent (e.g., \citealt{cal06}, \citealt{uza25}, and references therein). Furthermore, it has recently been proposed that varying constants (e.g., $m_{\rm e}(z)$, $\alpha(z)$) could provide a potential path towards alleviating the tension between Hubble constant values inferred from cosmic microwave background (CMB) observations and those derived from SNe\,Ia and Cepheid data in the context of the standard $\Lambda$-CDM cosmological model (e.g., \citealt{har20, sek21, lee23, sch25}).

In the last two decades, spectacular progress has been achieved in telescope sensitivity, which opened access to spectroscopic studies of very distant galaxies and quasars. This made it possible to compare $\alpha$ and $\mu$ in sources of different redshifts, $z$, and in different directions. The value of $\alpha$ can be derived by measuring the frequency of atomic or molecular fine-structure transitions, and the value of $\mu$ through that of molecular rotational, inversion, and torsion-rotational transitions, as long as at least two of these transitions have different sensitivity coefficients $K_{\alpha ,\mu}$ to variations of $\alpha$ and $\mu$, respectively.

So far, most spectroscopic studies of high-redshift objects have relied on observations of optical lines of atomic ions, detected in absorption toward distant quasars. They concern some 300 absorption systems with redshifts ranging from 0.1 to 6, i.e., with look-back times between 1 and 13 billion years, observed with the world's largest optical telescopes: Keck, VLT, and Subaru (for example, \citealt{web99,web01,mur03, kin12, eva14, mur17}). The analyses of those data yield rather discrepant conclusions regarding $\alpha$ (see, e.g., the review by \citealt{mar17}). Possible evidence of a variation of $\alpha$ with space-time was reported (\citealt{web99,mur01, web11, kin12}). The normalized difference between the values of $\alpha$ measured in remote quasars and that measured in a terrestrial laboratory, $\Delta \alpha / \alpha$, is of the order of a few parts per million, ppm. Furthermore, a 4-sigma dipole anisotropy with $\Delta \alpha / \alpha= 8 \pm 2$~ppm was obtained in favor of quasars located near the direction ${\rm R.A.} \simeq 17.5 {\rm h}, \rm{Dec.} \simeq -58^\circ$. This anisotropy is based on a first analysis of 128 archival absorption systems detected along the lines of sight to 68 quasars with the Keck/HIRES telescope \citep{mur03}, followed by a second analysis of 151 additional systems detected toward 60 quasars within the redshift range $0.5<z<3.5$ with the VLT/UVES \citep{web11, wil15}. Per contra, tighter limits to any variation of $\alpha$ were subsequently reported from dedicated, more sensitive observations carried out with the Subaru/HDS telescope ($\Delta \alpha / \alpha=3.0 \pm 2.8_{\rm stat}+2.0_{\rm syst}$ ppm, \citealt{mur17}), VLT/UVES ($\Delta \alpha / \alpha = 1.3 \pm 2.4_{\rm stat} + 1.0_{\rm syst}$ ppm, \citealt{mol13}), VLT/X-SHOOTER (up to $z=7.1$, \citealt{wil20}), and VLT/ESPRESSO ($\Delta \alpha / \alpha = 1.3 \pm 1.3_{\rm stat} + 0.4_{\rm syst}$ ppm, \citealt{mur22}). 

In any case and as pointed out by these and other works (see, e.g., \citealt{mur03,mur22, web11, whi15, wil15}), observations at optical wavelengths are hampered by the low sensitivity of UV atomic transitions to $\alpha$ variations and by the limited spectral resolving power of most optical spectrometers. The latter also suffer from systemic errors: i) echelle spectrometers are subject to physical distortions that are difficult to calibrate; this particularly affects the stacking of archival spectra, observed on different sources with different telescopes, that are averaged together to decrease the instrumental noise \citep{eva14}; ii) jitters of the quasar point-like images across the spectrometer slit, due to unstable atmosphere or inaccurate compensation of the sky rotation, induce wavelength measurement errors. Progress has recently been achieved through the installation of the new echelle spectrograph ESPRESSO on the VLT, specially designed for high resolution (resolving power $R=145,000$) and high spectral fidelity, in particular with the use of optical fibers, which removes environmental disturbances. So far, observations from only one remote quasar (HE0515$-$4414, $z_{\rm abs}=1.15$) with this spectrograph have been reported, yielding $\Delta \alpha / \alpha=1.3\pm 1.4_{\rm stat} + 0.4_{\rm syst}$ ppm or $|\Delta \alpha / \alpha| < 5.7$ ppm as $3\sigma$ upper limit \citep{mur22}. The same quasar holds the best constraint on any individual high-$z$ system, obtained by combining three independent measurements from three different spectrographs: $\Delta \alpha / \alpha= -0.62 \pm 0.50_{\rm stat} + 0.48_{\rm syst}$ or $|\Delta \alpha / \alpha| < 2.7$ ppm ($3\sigma$) \citep{mur22}.

It remains that the spectra of quasars at optical wavelengths are particularly rich in atomic lines from several species, and that the absorption profiles of those lines can be very broad. This leads to the blending of lines from different species with different $K_\alpha$ sensitivities. The resulting spectra are then difficult to analyze, even with advanced methods such as the Many Multiplet fitting \citep{web99}. In addition, the different isotopes of atoms have close, albeit not identical transition wavelengths, so that their lines partly overlap. Their relative abundances in high-$z$ absorbers are not known, and hence are usually set to their terrestrial value. This introduces unpredictable errors on the transition wavelength measurements, therefore, on $\Delta \alpha / \alpha$ (see, e.g., \citealt{web25}).

An alternative for measurements of $\alpha$ in remote galaxies consists of observing atomic, or mainly, molecular fine-structure lines at submillimeter wavelengths. This is now possible thanks to the high sensitivity and high angular and spectral resolution of current interferometers, such as ALMA and NOEMA. The advantages of submillimeter wave observations are multiple. On the technical side, the atmospheric dispersion effects are less critical at those wavelengths. The incoming sky signal, picked up by a horn, then channeled through waveguides to a digital spectrometer that is frequency-locked to an atomic clock, is more dependable and can be analyzed with a much higher frequency resolution than in any optical spectrometer (e.g., as high as $R \sim 10^8$ versus $\simeq 10^{4-5}$ for UVES and for ESPRESSO, the latest such instrument). On the target side, the submillimeter domain gives us access to molecular fine-structure and rotation lines. Molecular absorbers typically have narrower absorption profiles than the optical ones and show much less jammed spectra. In particular, the millimeter/submillimeter transitions of the different isotopologues of a given molecular species are well separated in frequency and their lines do not overlap.

In the present study, we focus on the two outstanding molecular absorbers at intermediate redshifts $z_{\rm abs}=0.89$ and $z_{\rm abs}=0.68$, which intercept the lines of sight to the radio-bright quasars PKS\,1830$-$211 and B\,0218$+$357, respectively. Both absorbers have already been subject to extended spectral studies at millimeter and submillimeter wavelengths and are known to be rich in molecular species (see, e.g., \citealt{wik95,wik96,wik98,mul06, mul11, mul14a, wal16}). In particular, the same constraint has been claimed for these two absorbers: $| \Delta \mu / \mu| < 3.6 \times 10^{-7}$ (at 3$\sigma$ confidence level ), using methanol lines toward PKS\,1830$-$211 on one hand (\citealt{mul21}, see also \citealt{bag13a,bag13b,kan15}), and using inversion lines of ammonia with respect to rotational lines of other species toward B\,0218$+$357 on the other hand \citep{mur08, kan11} \footnote{We note, however, that these upper limits were simply built as three times the final uncertainty of the measurement $x \pm \sigma_x$. Instead, taking the upper limits as $|x|+3\sigma_x$, we obtain $|\Delta \mu / \mu| < 0.54$~ppm toward PKS\,1830$-$211 and $|\Delta \mu / \mu| < 0.71$~ppm toward B\,0218$+$357, respectively.}. Interestingly, each quasar also lies near ($\lesssim 40^\circ$) one of the two poles of the $\alpha$ dipole anisotropy tentatively detected at $4\sigma$ by \cite{web11} in their analysis of the Keck and VLT data, making a good test of the dipole in the intermediate redshift range.

\section{Methylidyne and variations of $\alpha$ and $\mu$}

Methylidyne, CH, the very first molecule detected in the interstellar medium \citep{swi37} \footnote{While the 1937 detection of CH was done with the blue lines (A$^2\Delta$-X$^2\Pi$), the first radio detections of CH were done by \cite{ryd73} and \cite{tur74}.}, is known to be, with water, among the best proxies of molecular hydrogen \citep{ger16}. An open-shell radical, it has been extensively studied in the spectroscopic laboratory (see, e.g., \citealt{bro83, tru14}). 

\begin{figure}[h] \begin{center}
\includegraphics[width=8.8cm]{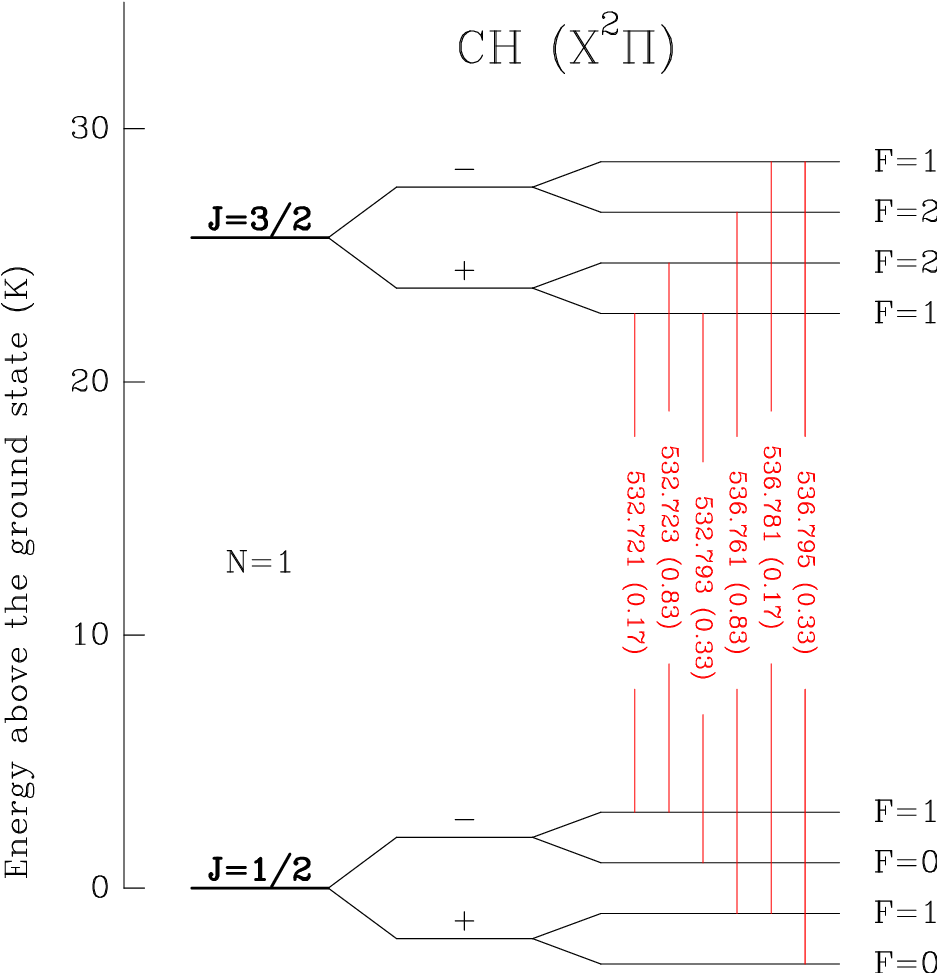}
\caption{Diagram of the lowest rotational energy levels of CH. The $\Lambda$-doublet and hyperfine levels are not to scale. The rest-frame frequencies of CH transitions observed by ALMA are indicated in red, in GHz, and their relative line strengths are given in parenthesis.}
\label{fig:CHdiag}
\end{center} \end{figure}

For CH, as for all diatomic molecules with an odd number of electrons, the interaction of molecular rotation with electronic motions leads to a partial uncoupling of the electron spin $S$ and the orbital angular momentum $L$ from the internuclear axis. Spin-uncoupling splits the CH $^2\Pi$ ground state into two substates $^2\Pi_{1/2}$ and $^2\Pi_{3/2}$ separated in energy by $A$, the spin-coupling constant. These substates correspond to the allowed orientations of the spin with respect to the angular momentum. $L$-uncoupling from the axis further splits the otherwise doubly degenerated rotational levels of each substate into two closely spaced sublevels denoted $+$ and $-$ according to their parity ($\Lambda$-doubling). Finally, the magnetic moment of the hydrogen nucleus further splits each of those sublevels into two hyperfine components labeled $F=1,0$ for $J=1/2$ and $F=1,2$ for $J=3/2$ (see, e.g.,\citealt{tru14} and Fig.\,\ref{fig:CHdiag}).

The rotational constant $B$ of light-weight molecules is large. By chance, for CH, its value ($B=425.5$~GHz) is close to half the value of $A$ ($A=834.8$~GHz), the separation between the spin substates. This has two beneficial consequences for us. Firstly, the $^2\Pi_{1/2}$ and $^2\Pi_{3/2}$ rotational level manifolds overlap in energy so that mixed transitions from one manifold to the other are allowed. In particular, transitions between the lowest rotational level of the $^2\Pi_{1/2}$ substate ($N=1$, $J=1/2$) and that of the $^2\Pi_{3/2}$ substate ($N=1$, $J=3/2$) are not only allowed, but have relatively large transition strengths (see Fig.\,\ref{fig:CHdiag} and Table~\ref{tab:spectro}). Secondly, their rest frequencies ($\sim 532.7$ and 536.8~GHz) are close to that of the $1_{10}-1_{01}$ ground rotational line of ortho-H$_2$O ($\sim 556.9$~GHz). All these three line sets can be observed almost continuously from $z \sim 0.1$ to $\sim 7$ within ALMA bands 2 to 8, and even simultaneously in favorable cases, at $z = 0.68$ and 0.89 for example.

Pure rotational and mixed transitions have quite different sensitivities to variations of $\alpha$ and $\mu$. The former are sensitive to $\mu$, while the latter are essentially sensitive to $\alpha$, so that their combination provides a good test of the variations of these two constants. The sensitivity coefficients $K_\alpha$ and $K_\mu$ of the CH mixed transitions mentioned above have been calculated by \cite{nij12} (see Table~\ref{tab:spectro}). They find $K_\alpha = 1.59, 1.57$ and $K_\mu = -0.20, -0.22$ for the $+\rightarrow -$ ($\sim 533$~GHz) and $-\rightarrow +$ ($\sim 537$~GHz) transitions, respectively. The sensitivity differences between those CH lines and a pure rotational line, such as the $1_{10}-1_{01}$ ground-state transition of ortho-H$_2$O, with $K_\alpha = 0$ and $K_\mu = -1$, are then $\Delta K_\alpha = 1.6$ and $\Delta K_\mu = 0.8$. Any variations $\Delta \alpha$ and $\Delta \mu$ would therefore introduce a velocity shift $\delta v$ between the spectra of CH and H$_2$O:
\begin{equation} \label{eq:df/f}
\frac{\delta v}{c} = 1.6 \times \frac{\Delta \alpha}{\alpha} + 0.8 \times \frac{\Delta \mu}{\mu}.
\end{equation}
\noindent In particular, the sensitivity $\Delta K_\alpha$ is more than one order of magnitude higher than that governing the UV lines of metal ions (see, e.g., Fig.\,3 by \citealt{mur22}) which until now constitute the bulk of previous high-redshift constraints on $\Delta \alpha / \alpha$. Equation\,\ref{eq:df/f} shows that the variations of $\alpha$ and $\mu$ cannot be disentangled. However, since very low upper limits to the variation of $\mu$ ($|\Delta \mu / \mu| \leq$ few $10^{-7}$) have already been independently derived for our molecular absorbers (see the end of Sec.\,\ref{sec:intro}), the observation of CH and H$_2$O lines provides an exceptionally sensitive test of the variations of $\alpha$.

We note that the  $J=1/2^+-1/2^-$ (3.3~GHz, rest) and $J=3/2^+-3/2^-$ (0.7~GHz, rest) $\Lambda$-doublets have been previously used (in conjunction with OH 1.6~GHz lines) to constrain local variations of $\mu$ and $\alpha$ in sources of the Milky Way by \cite{tru13} \footnote{They find $\Delta \alpha / \alpha = (0.3 \pm 1.1) \times 10^{-7}$ and $\Delta \mu / \mu = (-0.7 \pm 2.2) \times 10^{-7}$, assuming that there is either a change in $\alpha$ or $\mu$, but not in both.}. However, those radio lines typically have optical depths several orders of magnitude lower than those of the submillimeter ones at 533--537~GHz in the interstellar medium.

\begin{table*}[ht] 
\caption{Spectroscopic parameters of the CH and H$_2$O transitions considered here and their sensitivity coefficients to variations of $\mu$ and $\alpha$.}
\label{tab:spectro}
\begin{center} \begin{tabular}{cccccccc}
\hline
Species & Line & Rest Freq. $^{(a)}$ & Sky Freq. $^{(b)}$         & $S_{ul}$ $^{(c)}$ & $E_l$ $^{(d)}$ & $K_\mu$ $^{(e)}$ & $K_\alpha$ $^{(e)}$ \\
        &      & (kHz)       & (GHz) &          & (K)  &        & \\
\hline
CH & $\Omega$'=1/2-3/2, J=1/2-3/2 F=1-1 & 532,721,588.6 (0.6) & 282.488 & 0.17 & 0.2 & $-0.20$ & +1.59 \\
CH & $\Omega$'=1/2-3/2, J=1/2-3/2 F=1-2 & 532,723,889.3 (0.6) & 282.489 & 0.83 & 0.2 & $-0.20$ & +1.59 \\
CH & $\Omega$'=1/2-3/2, J=1/2-3/2 F=0-1 & 532,793,274.6 (0.6) & 282.526 & 0.33 & 0.2 & $-0.20$ & +1.59 \\
CH & $\Omega$'=1/2-3/2, J=1/2-3/2 F=1-2 & 536,761,046.3 (0.6) & 284.630 & 0.83 & 0.0 & $-0.22$ & +1.57 \\
CH & $\Omega$'=1/2-3/2, J=1/2-3/2 F=1-1 & 536,781,856.3 (0.6) & 284.641 & 0.17 & 0.0 & $-0.22$ & +1.57 \\
CH & $\Omega$'=1/2-3/2, J=1/2-3/2 F=0-1 & 536,795,569.5 (0.6) & 284.648 & 0.33 & 0.0 & $-0.22$ & +1.57 \\
H$_2$O  & J,K$_{\rm a}$,K$_{\rm c}$=1$_{10}$--1$_{01}$ (ortho) & 556,935,987.7 (0.3) & 295.328 & 4.5 & 0.0 & $-1$ & 0 \\
H$_2^{18}$O  & J,K$_{\rm a}$,K$_{\rm c}$=1$_{10}$--1$_{01}$ (ortho) & 547,676,470 (15)  & 290.418 & 4.5 & 0.0 & $-1$ & 0 \\
\hline
\end{tabular}
\tablefoot{
(a) For CH, frequencies are taken from \cite{tru14}. For H$_2$O, the frequency is taken from \cite{caz09}. The hyperfine structure of the ortho-H$_2$O 1$_{10}$--1$_{01}$ line spreads over less than a few tens of m\,s$^{-1}$ and is thus negligible in our work. For H$_2^{18}$O, the frequency is taken from \cite{gol06}.
(b) Sky frequencies are calculated for $z$=0.88582.
(c) Line strength.
(d) Energy of the lower level.
(e) The sensitivity coefficients $K_\mu$ and $K_\alpha$ for CH are taken from \cite{nij12}.
}
\end{center} \end{table*}

\section{Observations}

\begin{figure*}[ht!] \begin{center}
\includegraphics[width=\textwidth]{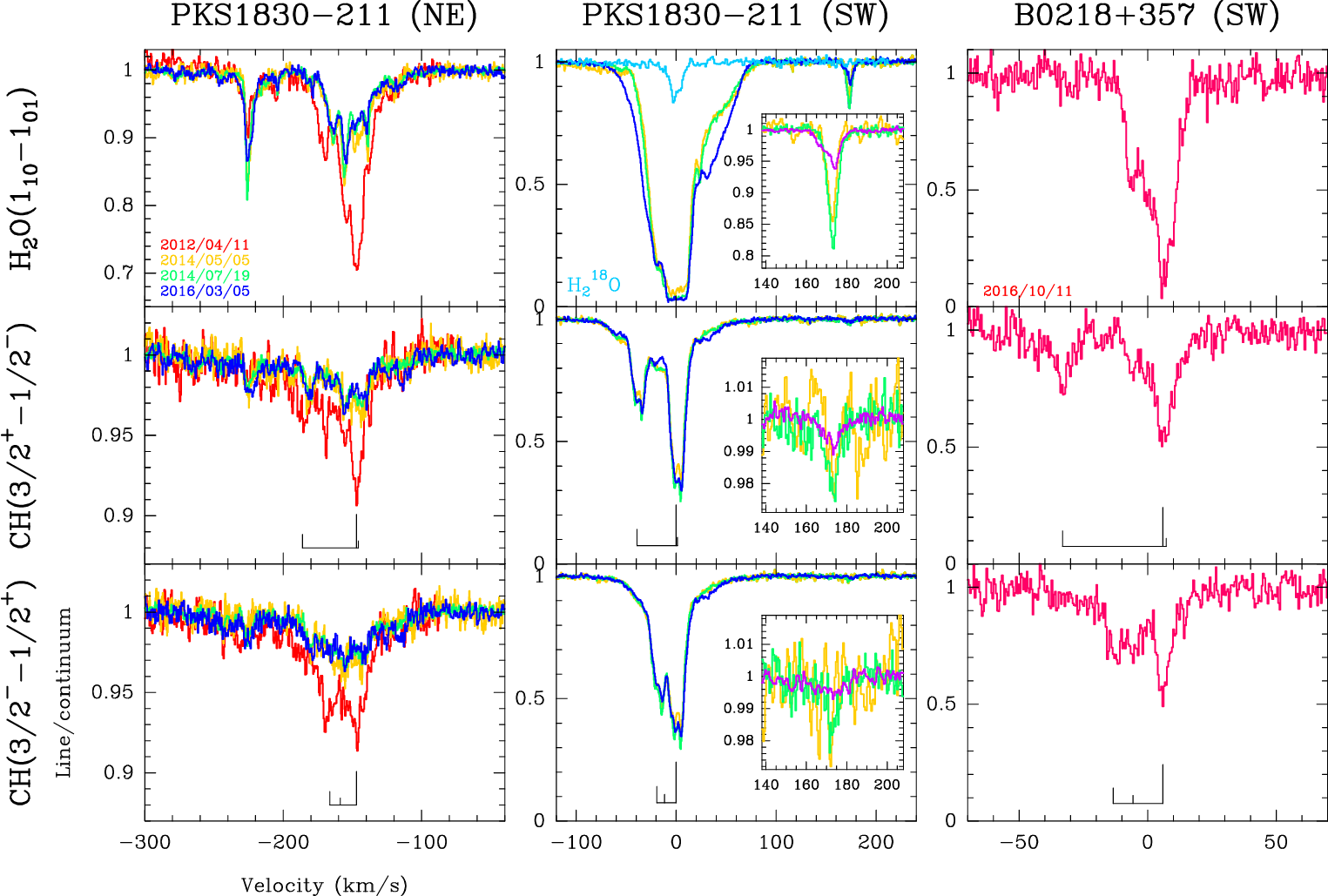}
\caption{Spectra of CH and H$_2$O lines toward PKS\,1830$-$211 NE and SW images (left, middle, respectively), and toward B\,0218$+$357(SW) (right) observed by ALMA at different epochs. The spectra are colored by dates, as indicated in the top-left and top-right boxes. The hyperfine structure of each CH $\Lambda$-doublet is indicated at the bottom of spectra. The spectrum of the H$_2^{18}$O isotopologue, taken on 2014/05/05, is also shown for PKS\,1830$-$211(SW) (top-middle box, in light blue). A zoom on the weak $v \sim +170$~km\,s$^{-1}$ velocity component is shown for PKS\,1830$-$211(SW), with the 2016 averaged spectra (in purple) instead of that taken on 2016/03/05.}
\label{fig:allfit}
\end{center} \end{figure*}

\begin{figure}[ht!] \begin{center}
\includegraphics[width=8.8cm]{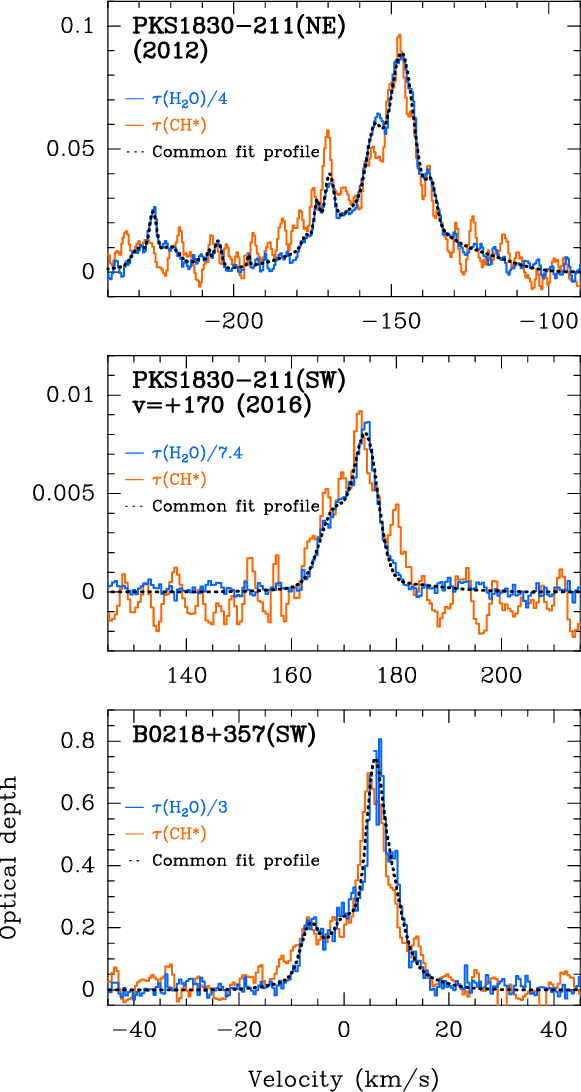}
\caption{Overlay of the opacity profiles of CH (average of the two $\Lambda$-doublets, deconvolved from their respective hyperfine structure) and H$_2$O (scaled by the opacity ratio $\gamma_\tau$), on top of their common fit profile (see Sec.\,\ref{sec:gaussfit}) for 
PKS\,1830$-$211(NE) (top, 2012 data), the isolated velocity component near $v \sim +170$~km\,s$^{-1}$ toward PKS\,1830$-$211(SW) (middle, average of the 2016 data), and B\,0218$+$357(SW) (bottom). Note the span of two orders of magnitude between the optical depths of these features.}
\label{fig:specdeconv}
\end{center} \end{figure}

\subsection{PKS\,1830$-$211}

The spectra of CH and H$_2$O were observed simultaneously with ALMA at several epochs between 2012 and 2016. One visit was made on 11 April 2012 during ALMA Cycle 0 Early Science, two visits were made in 2014 (5 May and 17 July), and a monitoring consisting of multiple visits spread over six months and with an irregular cadence between 6-21~days was performed in 2016. Those data have previously been presented by \cite{mul14a,mul23}. We refer to these publications for the complete descriptions of the data reduction and summarize here only the relevant information for this work.

The number of 12\,m antennas in the array was 16 for 2012 Cycle~0 observations, between 31 and 33 in 2014, and between 36 and 45 during the 2016 monitoring. The on-source integration time per visit was 50~min in 2012, $\sim 30$~min in 2014, and 7.5~min/visit during the 2016 campaign. The resulting synthesized beam was $\lesssim 0.5$\arcsec\ for all observations, allowing us to well resolve the two lensed images of the quasar, separated by $\sim 1$\arcsec\ (e.g., \citealt{sub90,fry97,mul20}). The amount of precipitable water vapor was typically between 0.6 and 2~mm. The correlator was set up with separate spectral windows to cover the CH and H$_2$O lines, redshifted in the frequency range of ALMA Band~7 receivers \citep{mah12}, with a velocity resolution of 1.0~km\,s$^{-1}$ (2012 data) and 0.6~km\,s$^{-1}$ (2014-2016 data). Data reduction was performed in CASA\footnote{\url{http://casa.nrao.edu/}} with a standard calibration procedure. The visibilities of PKS\,1830$-$211 were self-calibrated and the spectra were extracted from visibility fitting using the Python tool UVMultiFit (\citealt{mar14}), taking a model consisting of two point sources with free amplitudes and relative positions fixed at the known locations of the two lensed images of the quasar (hereafter NE and SW images). The final spectra (Fig.\ref{fig:allfit}, left, and middle columns) were normalized to the continuum level of each image.

On 5 May 2014, a separate tuning including the ground-state transition of the H$_2^{18}$O isotopologue was observed for $\sim 36$~min on-source, providing a clear detection toward PKS\,1830$-$211(SW) (Fig.\ref{fig:allfit}, top-middle box).

\subsection{B\,0218$+$357}

Both CH and H$_2$O were observed simultaneously with ALMA on 11 October 2016 with the Band~7 receivers. The synthesized beam was $\sim 0.1$\arcsec, smaller than the separation of 0.33\arcsec\ between the two lensed images of the quasar \citep{ode92,pat93}. The correlator was setup with spectral windows of 0.938~GHz width and a velocity resolution of 0.5~km\,s$^{-1}$. The calibration of the data and the extraction of the spectra with UVMultiFit followed the same procedure as described for PKS\,1830$-$211. The spectra toward B\,0218$+$357 (Fig.\ref{fig:allfit}, right column) are noticeably noisier than those toward PKS\,1830$-$211, due to its lower flux density, its lower elevation from the ALMA site ($\sim 30^\circ$ maximum elevation, whereas PKS\,1830$-$211 transits near zenith), and the limited integration time ($\sim 15$~min on-source, with 44 antennas in the array and a low amount of precipitable water vapor of $\sim 0.5$~mm during the observations).

\section{Analysis and results}

\subsection{Description of spectra}

Both absorption systems, PKS\,1830$-$211 and B\,0218$+$357, share very similar properties, with the background quasar lensed by the foreground absorber in two main bright and compact images. 

In the case of PKS\,1830$-$211, both lines of sight present molecular absorption \citep{wik98,mul06,mul14a}. For PKS\,1830$-$211(NE), the absorption profile is composed of a series of narrow (a few km\,s$^{-1}$ wide) and optically thin ($\tau \lesssim 0.1$ for H$_2$O) features between $v \sim -250$ and $-100$~km\,s$^{-1}$ (Fig.\,\ref{fig:allfit}, left column), well adapted for the comparison of CH and H$_2$O profiles. For PKS\,1830$-$211(SW) (Fig.\,\ref{fig:allfit}, middle column), there is a main absorption feature at $v=0$~km\,s$^{-1}$, with a width of 10--20~km\,s$^{-1}$ for optically thin lines (e.g., CH, H$_2^{18}$O). However, this central component becomes saturated and flat bottomed over a range of $\sim 20$~km\,s$^{-1}$ for the very optically thick line of H$_2$O. This region is therefore not useful for our purpose of comparing the kinematics of CH and H$_2$O directly. On the other hand, the broad wings (covering up to $\sim 150$~km\,s$^{-1}$) seen on the H$_2$O absorption quickly disappear below the noise level for CH. Therefore, it is difficult to make a tight comparison of the CH and H$_2$O profile for the main feature of PKS\,1830$-$211(SW), unless the optically thin H$_2^{18}$O isotopologue is used in complement (as was done for the 2014/05/05 data). While several other weak velocity components were identified \citep{mul11,mul14a}, we highlight the presence of the narrow (a few km\,s$^{-1}$ wide) and isolated feature near $v=+170$~km\,s$^{-1}$, which can then be used to constrain the kinematics of CH versus H$_2$O, although with a limited signal-to-noise ratio. Interestingly, the quasar activity, with intrinsic changes in its morphology effectively amplified by the lens geometry, introduces a time variability of the absorption profile with a time scale of a few days to months \citep{mul08,sch15,mul23}, which, in practice, is equivalent to multiplying the effective number of lines of sight from different absorber systems.

In the case of B\,0218$+$357, only the line of sight to the southwestern lensed image shows molecular absorption \citep{men96,mul07}. The absorption profile resembles that of PKS\,1830$-$211(SW), with a handful of velocity components within $\lesssim 30$~km\,s$^{-1}$ (Fig.\,\ref{fig:allfit}, right column). The H$_2$O absorption does not reach saturation, while the peak opacity of CH appears to be intermediate between those of PKS\,1830$-$211 NE and SW profiles. 

In the following, we present our analysis method and results for each line of sight, as well as tests on synthetic spectra.

\subsection{Fitting of a common multi-Gaussian profile} \label{sec:gaussfit}

We assume that the spectra of both CH and H$_2$O can be described by a common opacity profile $G_\tau(v)$, composed of the sum of individual Gaussian components. The normalized absorption spectra are then:
\begin{equation}
I_{\rm abs}(v)  = 1 - f_c \times (1-{\rm e}^{-G_\tau(v)}),
\end{equation}
where $f_c$ is the continuum covering factor ($0 \leq f_c \leq 1 $). During the fit, the profile $G_\tau(v)$ was normalized to a fiducial CH hyperfine component of intensity 1 (hereafter, CH$^*$), and convolved to the hyperfine structure of CH, with the relative intensities of hyperfine components following their statistical weights (see Table~\ref{tab:spectro} and discussion in Sec.\,\ref{sec:spectroadv}). For H$_2$O, the opacity profile $G_\tau(v)$ was scaled by a factor $\gamma_\tau = \tau(\rm H_2O)/\tau(\rm CH^*)$. Lastly, we introduced a relative bulk velocity offset $\delta v$ (that is, given to each individual velocity component of the opacity profile $G_\tau$) between the two species. Therefore, the fitting procedure included the following parameters: centroid velocity, linewidth, and peak opacity for each Gaussian component, plus the covering factor, if not fixed, the global scaling ratio $\gamma_\tau$, and the bulk velocity offset $\delta v$.

For the fit, we used the Python scipy.curve\_fit least-squares routine. The uncertainties were derived as the square root of the diagonal elements of the covariance matrix multiplied by the reduced $\chi^2$. The number of Gaussian velocity components was adjusted by successive trials, until the residuals became free from features above the noise level and the reduced $\chi^2$ close to unity. We ended up fixing the number of Gaussians to 17 for the line of sight toward PKS\,1830$-$211(NE), 11 for PKS\,1830$-$211(SW), 3 for the $v=+170$~km\,s$^{-1}$ isolated velocity component toward PKS\,1830$-$211(SW), and 5 for B\,0218$+$357(SW). The same fitting setups were used for our tests with synthetic spectra (see \S\,\ref{sec:synthetic_spec}). 

The best-fit results for $\delta v$ and $\gamma_\tau$ are given in Table~\ref{tab:velooff} together with the $\chi^2$ values for the three lines of sight. All reduced $\chi^2$ are close to unity, suggesting that the CH and H$_2$O spectra are reasonably well represented by our assumption of a common opacity profile. To illustrate this, we have overlaid the latter on top of the opacity profiles of CH and H$_2$O for several absorption features in Fig.\,\ref{fig:specdeconv}. For CH, we have deconvolved each CH $\Lambda$-doublet from its hyperfine structure following the iterative CLEAN algorithm \citep{hog74}, with the relative intensities of the hyperfine components fixed to their statistical equilibrium values. The two profiles were normalized to an equivalent single hyperfine component of intensity 1 and averaged together. The opacity profile of H$_2$O was scaled by the corresponding ratio $\gamma_\tau$. The fit profiles are clearly driven by the higher signal-to-noise ratio of the H$_2$O line but are remarkably consistent with the CH profile, within uncertainties.

\begin{table}[ht] 
\caption{Bulk velocity offsets ($\delta v$) and opacity scaling ratios ($\gamma_\tau$) between CH and H$_2$O spectra toward PKS\,1830$-$211 and B\,0218$+$357 from fitting a common opacity profile to both species (including H$_2^{18}$O when specified).}
\label{tab:velooff}
\begin{center} \begin{tabular}{cccc}
\hline
Date & $\delta v$ & $\gamma_\tau$ & $\chi^2$ \\
     & (km\,s$^{-1}$)     &                 &    \\
\hline
\multicolumn{4}{c}{PKS\,1830$-$211(NE)} \\
\hline
2012/04/11 & $ 0.04 \pm 0.13 $ & $3.97 \pm 0.05$ & 1.19 \\
2014/05/05 & $-0.21 \pm 0.08 $ & $4.47 \pm 0.06$ & 1.22 \\
2014/07/19 & $ 0.03 \pm 0.06 $ & $4.49 \pm 0.06$ & 1.88 \\
2016/03/05 & $-0.30 \pm 0.11 $ & $3.95 \pm 0.06$ & 1.46 \\
2016/03/26 & $ 0.65 \pm 0.14 $ & $3.83 \pm 0.06$ & 1.52 \\
2016/04/03 & $ 0.07 \pm 0.15 $ & $3.93\pm  0.07$ & 1.21 \\
2016/04/10 & $ 0.08 \pm 0.11 $ & $3.89 \pm 0.06$ & 1.15 \\
2016/04/16 & $-0.32 \pm 0.14 $ & $3.77 \pm 0.07$ & 1.13 \\
2016/04/23 & $-0.02 \pm 0.16 $ & $3.84 \pm 0.07$ & 1.18 \\
2016/05/02 & $ 0.25 \pm 0.19 $ & $3.61 \pm 0.07$ & 1.07 \\
2016/05/08 & $ 0.31 \pm 0.14 $ & $4.00 \pm 0.09$ & 1.19 \\
2016/05/17 & $ 0.08 \pm 0.15 $ & $4.04 \pm 0.08$ & 1.08 \\
2016/06/15 & $-0.40 \pm 0.14 $ & $4.01 \pm 0.08$ & 1.16 \\
2016/06/22 & $-0.22 \pm 0.19 $ & $4.06 \pm 0.10$ & 1.05 \\
2016/07/01 & $-0.24 \pm 0.15 $ & $4.14 \pm 0.09$ & 1.06 \\
2016/07/14 & $-0.68 \pm 0.17 $ & $3.87 \pm 0.11$ & 1.12 \\
2016/07/28 & $ 0.01 \pm 0.15 $ & $3.91 \pm 0.07$ & 1.12 \\
2016/09/08 & $-0.03 \pm 0.20 $ & $4.05 \pm 0.13$ & 0.96 \\
Weighted average: & $-0.048 \pm 0.028$ & $4.00 \pm 0.02$ & \\
\hline
\multicolumn{4}{c}{PKS\,1830$-$211(SW) $v \sim 0$ km\,s$^{-1}$, including H$_2^{18}$O} \\
\hline
2014/05/05 & $0.041 \pm 0.070$ & $8.33 \pm 0.08$ & 1.40 \\
\hline
\multicolumn{4}{c}{PKS\,1830$-$211(SW) $v=+170$~km\,s$^{-1}$} \\
\hline
2014/05/05 & $0.12 \pm 0.39$ & $8.3 \pm 1.1$ & 1.08\\
2014/07/19 & $-0.31 \pm 0.17$ & $8.6 \pm 0.4$ & 1.21 \\
2016 (averaged) & $ 0.13 \pm 0.21 $ & $7.4 \pm 0.3$ & 1.10 \\
Weighted average: & $-0.11 \pm 0.13$ & $7.9 \pm 0.2$ & \\
\hline
\multicolumn{4}{c}{B\,0218$+$357(SW)} \\
\hline
2016/10/11 & $ -0.13 \pm 0.14 $ & $3.01 \pm 0.10$ & 1.07 \\ 
\hline
\end{tabular} \end{center} \end{table}

\subsection{Results for the PKS\,1830$-$211(NE) sightline}

Since all velocity offset measurements have been made from different datasets, taken at different epochs and with clear time variations of the absorption profile \citep{mul23}, we consider them statistically independent. Taking their weighted average for all individual epochs, we obtain a value $\delta v = -0.048 \pm 0.028$~km\,s$^{-1}$. For the opacity scaling ratio between H$_2$O and CH, we obtain an average $\gamma_\tau=4.00 \pm 0.02$, which is consistent with the value that we would expect taking the relative abundances of CH and H$_2$O observed along the Milky Way's diffuse sightlines (\citealt{she08,fla13}, see also the discussion by \citealt{mul23} on the correlation of CH and H$_2$O opacities, where they estimate $\gamma_\tau$ ranging between $\sim 3$ and $\sim 9$). A fixed covering factor $f_c = 1$ was assumed for this line of sight, since the lines have low apparent optical depths and the profile does not appear to be saturated. In fact, \cite{mul17} found a covering factor $f_c > 0.8$ for this line of sight. Changing the value of $f_c$ to 0.8 in our fits yielded similar velocity offsets and scaling ratios within uncertainties. As long as the lines are optically thin, we expect a degeneracy between $f_c$ and the opacities of the individual velocity components, thus absorbed in the fit.

Our method of fitting a common profile may introduce some systematics if the absorption profiles of CH and H$_2$O are not strictly identical, i.e. if the scaling ratios $\gamma_\tau$ change between individual velocity components. However, we argue that both the number of independent velocity components in the profile and the number of epochs (here the time variability of the absorption profile between different epochs helps us to improve the statistics) tend to minimize any systematic effect between CH and H$_2$O. This is confirmed by testing the effect of varying the opacity ratio $\gamma_\tau$ for each velocity component on synthetic spectra, see Sec.\,\ref{sec:synthetic_spec} below.

\subsection{Results for PKS\,1830$-$211(SW)}

For the epoch when we have H$_2^{18}$O observed within the same day as CH and H$_2$O, we obtain a bulk velocity offset $\delta v = 0.041 \pm 0.070$~km\,s$^{-1}$ between CH and the water isotopologues, consistent with no kinematics difference. Regarding its uncertainty, this measurement is at the level of the best value of a single visit obtained toward PKS\,1830$-$211(NE), which suggests that the very high signal-to-noise ratio of the SW spectra just compensates for the larger number and narrowness of the velocity components in the NE line of sight. The opacity ratio between H$_2$O and CH is $\gamma_\tau = 8.33 \pm 0.08$, and we find an isotopologue abundance ratio H$_2^{16}$O/H$_2^{18}$O = $67 \pm 2$, consistent with previous measurements \citep{mul23}. The covering factor is well determined by the saturated part of the water line and was kept as a free parameter of the fit, resulting in $f_c = 0.944 \pm 0.002$.

In addition, we obtain three more measurements using the isolated $v=170$~km\,s$^{-1}$ component alone, although with somewhat large uncertainties. All spectra obtained during the 2016 monitoring were averaged together before fit, in order to improve the signal-to-noise ratio. Combining these three measurements, we obtain a value $\delta v = -0.11 \pm 0.13$~km\,s$^{-1}$. The opacity ratio $\gamma_\tau$ is similar between this component and the main feature near $v=0$~km\,s$^{-1}$, and a factor two higher than that obtained for PKS\,1830$-$211(NE), which may reflect a different physico-chemical environment between the two lines of sight, as previously suggested by \cite{mul23}.

Combining the results from the $v=0$ and $v=170$~km\,s$^{-1}$ features, we obtain $\delta v = 0.005 \pm 0.061$~km\,s$^{-1}$.

\subsection{Results for B\,0218$+$357(SW)}

The single-epoch measurement of the velocity offset between CH and H$_2$O, $\delta v = -0.13 \pm 0.14$~km\,s$^{-1}$, is consistent with zero and its uncertainty is comparable to that of individual measurements toward PKS\,1830$-$211(NE). Because the signal-to-noise ratio of the spectra is limited, and with the lack of additional constraints on the covering factor $f_c$, we have explored the effect of varying $f_c$ between 0.8 and 1.0, by fixing its value during the fit. There was no significant impact on $\delta v$ within uncertainties. On the other hand, the opacity scaling ratio depends on $f_c$, with values of $\gamma_\tau$ increasing from $3.01 \pm 0.10$, for $f_c= 1.0$, to $3.57 \pm 0.15$, for $f_c= 0.8$. This is explained because the absorption has a higher optical depth than in the case of PKS\,1830$-$211(NE). The values of the scaling ratios are again consistent with the range of values observed in diffuse sightlines in the Milky Way. They are closer to those obtained in the NE line of sight of PKS\,1830$-$211 than those in its SW line of sight, adding to previous evidence collected by \cite{wal16,wal19} that the absorbing material in B\,0218$+$357(SW) have properties close to PKS\,1830$-$211(NE).

\subsection{Test on synthetic spectra} \label{sec:synthetic_spec}

In order to test the robustness of our fitting procedure, we have run Monte Carlo simulations on synthetic spectra. The initial opacity profile was composed of 13 Gaussian components for PKS\,1830$-$211(NE), and 5 components for PKS\,1830$-$211(SW), with properties similar to those of the real spectra observed in 2016. A fixed bulk velocity offset $\delta v$=1~km\,s$^{-1}$ and scaling ratios $\gamma_{\tau}$=4.0 and 8.0 were set between the two species, for the NE and SW sightlines, respectively. The covering factor $f_c$ was set to 1.0 and 0.95 for the NE and SW spectra, respectively. For each Monte Carlo run, we added a Gaussian channel noise of rms $\sigma=0.5\%$, similar to the noise of our observed spectra \citep{mul23}. We further let the amplitude of each Gaussian velocity component of the H$_2$O profile vary by the factor 1+$f(\lambda)$ where $f(\lambda)$ is a random number taken from a Gaussian probability distribution of dispersion $\lambda$. This allows us to test the effect of a possible varying opacity ratio between CH and H$_2$O. The fitting procedure was then strictly identical to the treatment of real observational data, as described in \S\,\ref{sec:gaussfit}. All this process was iterated several hundred times to extract the statistical values shown in Table~\ref{tab:synthetic_spec}.

These simulations suggest that the bulk velocity offset can be recovered with typical uncertainties similar to the real observations (Table~\ref{tab:velooff}) even in the case of variation of up to 20--30\% for PKS\,1830$-$211(NE). In contrast, the fit quality of PKS\,1830$-$211(SW) spectra depends much more critically on the amount of amplitude variations (i.e., the spread in the correlation between CH and H$_2$O). For them, the fit uncertainties of $\delta v$ jump to values higher than 0.3~km\,s$^{-1}$, as soon as 10\% variations are introduced. This is due to both the smaller number of velocity components (5 versus 13 for PKS\,1830$-$211(NE)) and to the saturation of H$_2$O, where the fitting loses constraints on the bulk velocity offset for H$_2$O saturated velocity components. The covering factor $f_c$, set as a free parameter for the fit of PKS\,1830$-$211(SW) spectra, was always recovered with uncertainties lower than a few per mil.

In either case, we note that the uncertainties of the scaling ratio $\gamma_\tau$ grow faster than those of the velocity offset when the amplitude variations per individual velocity components increase. The relatively small dispersion of the average $\gamma_\tau$ from our real PKS\,1830$-$211(NE) spectra in Table~\ref{tab:velooff} is then a good indication that the method of fitting a common opacity profile is robust, at least for this line of sight.

\begin{table}[ht] 
\caption{Fit results of Monte Carlo simulations of CH and H$_2$O synthetic spectra with random amplitude variations governed by the dispersion $\lambda$. We set a fixed bulk velocity offset $\delta v = 1.0$~km\,s$^{-1}$ and the scaling ratios $\gamma_\tau=4.0$, for PKS\,1830$-$211(NE) and $\gamma_\tau=8.0$, for PKS\,1830$-$211(SW).}
\label{tab:synthetic_spec}

\begin{center} \begin{tabular}{cccc}
\hline
$\lambda$ & $\langle \delta v \rangle$ & $\langle \gamma_\tau \rangle$ & $\langle \chi^2 \rangle$ \\
\hline
\multicolumn{4}{c}{PKS\,1830$-$211(NE)} \\
\hline
0.0 & $1.00 \pm 0.09$ & $4.00 \pm 0.05$ & $1.01 \pm 0.06$ \\
0.1 & $1.01 \pm 0.10$ & $4.00 \pm 0.17$ & $1.02\pm 0.05$ \\
0.2 & $1.00 \pm 0.12$ & $4.08 \pm 0.36$ & $1.10 \pm 0.09$ \\
0.3 & $1.02 \pm 0.17$ & $4.13 \pm 0.48$ & $1.17 \pm 0.14$ \\
0.5 & $1.02 \pm 0.54$ & $4.30 \pm 0.73$ & $1.43 \pm 0.31$ \\
\hline
\multicolumn{4}{c}{PKS\,1830$-$211(SW)} \\
\hline
0.0 & $1.01 \pm 0.04$ & $8.02 \pm 0.04$ & $1.37 \pm 0.45$ \\
0.1 & $0.98 \pm 0.30$ & $7.98 \pm 0.50$ & $1.45 \pm 0.47$ \\
0.2 & $1.03 \pm 0.55$ & $8.03 \pm 1.08$ & $1.72 \pm 0.71$ \\
0.3 & $1.06 \pm 0.80$ & $8.10 \pm 1.50$ & $2.20 \pm 1.32$ \\
\hline

\end{tabular}
\tablefoot{$\lambda$ represents the dispersion of amplitude variation introduced on each velocity component of the H$_2$O spectra, see \S\,\ref{sec:synthetic_spec}.}
\end{center} \end{table}

\section{Discussion} \label{sec:discussion}

We can convert the final weighted average measurement $\delta v  = -0.048 \pm 0.028$~km\,s$^{-1}$ between CH and H$_2$O toward PKS\,1830$-$211(NE) into constraints on the variations of $\mu$ and $\alpha$ using Eq.\,\ref{eq:df/f}. All the upper limits below are built as $|x|+ 3\sigma_x$. If we assume that there are no variations of $\mu$, we obtain $|\Delta \alpha / \alpha| < 0.28$~ppm. If we now assume that there are no variations of $\alpha$, we get $|\Delta \mu / \mu| < 0.55$~ppm. This last constraint is similar to that obtained by \cite{mul21}, $|\Delta \mu / \mu| < 0.54$~ppm, using methanol lines along the line of sight to PKS\,1830$-$211(SW). Combining this independent constraint on $\Delta \mu / \mu$ from methanol and our new constraint from CH--H$_2$O lines, we derive $|\Delta \alpha / \alpha| < 0.55$~ppm, which is the tightest cosmological constraint on $\Delta \alpha / \alpha$ so far, a factor four deeper than for any other single high-$z$ system (see, e.g., \citealt{mur22} and a compilation by \citealt{uza25}).

Toward B\,0218$+$357(SW), we take our measurement $\delta v = -0.13 \pm 0.14$~km\,s$^{-1}$ and the independent constraint $|\Delta \mu / \mu|<7.1$~ppm from \cite{kan11} and obtain the final constraint $|\Delta \alpha / \alpha| < 1.5$~ppm.

Our measurements provide a radio alternative to optical studies to test the apparent $\alpha$ dipole anisotropy described in the introduction. The direction to PKS\,1830$-$211 is $\Theta \sim 39^\circ$ away from the pole in the southern hemisphere (R.A. $\simeq 17.5$~h; Dec.~$\simeq -58^\circ$), and the direction to B\,0218$+$357 is $\Theta \sim 38^\circ$ away from the opposite pole. With a dipole of $\Delta \alpha / \alpha$ described as $\mathcal{D}\cos{(\Theta)}$, we find that the difference in variation $\Delta \alpha / \alpha$ between the directions to PKS\,1830$-$211 and B\,0218$+$357 would then be $\sim 1.6 \times \mathcal{D}$ (i.e., $\sim 13 $~ppm). Taking the bulk velocity offsets between CH and H$_2$O toward PKS\,1830$-$211(NE) and B\,0218$+$357(SW) and the corresponding independent upper limits of $\Delta \mu / \mu$, as mentioned above, we derive an upper limit of 1.6~ppm (at 3$\sigma$) for the difference in $\Delta \alpha / \alpha$ between our two absorbers. Our measurements, although at different redshifts and based on only two sources, would then rule out an amplitude $\mathcal{D} \gtrsim 1$~ppm ($3\sigma$) for the dipole, in contradiction with the past value measured from atomic absorbers (e.g., \citealt{web11}). Thereby, our radio measurements add up to the subsequent optical studies examining various systematics to rule out the $\alpha$ dipole anisotropy model.

\subsection{Robustness of our constraints} 

The stringency and robustness of our constraints on $\Delta \alpha / \alpha$ are ensured by the combination of several factors, including the spectroscopic properties of CH and H$_2$O lines, the properties of the absorber systems PKS\,1830$-$211 and B\,0218$+$357, and our observational setup. We discuss these factors in the following subsections.

\subsubsection{Spectroscopy} \label{sec:spectroadv}

As already highlighted, the CH $J$=3/2-1/2 and H$_2$O $1_{10}$-$1_{01}$ transitions are particularly sensitive to variations of $\alpha$ and $\mu$, with a difference in the sensitivity coefficients $\Delta K_{\alpha} = 1.6$ and $\Delta K_{\mu} = 0.8$ between them. All lines are close in frequency and can be observed simultaneously with ALMA at $z=0.89$ and 0.68 within the same tuning (although with the current system, H$_2^{18}$O must be observed with a different tuning).

The rest frequencies of the submillimeter transitions of CH and H$_2$O have been determined with high precision (Table~\ref{tab:spectro} and, e.g., \citealt{tru14} for CH, and \citealt{caz09, alt25}, for H$_2$O), with uncertainties lower than 1\,kHz, i.e. $< 1$~m\,s$^{-1}$ when converted to velocity. This is two to three orders of magnitude lower than the widths of the velocity components and than the fit uncertainties of the velocity offsets $\delta v$, and therefore negligible in our final error budget.

Moreover, and unlike for atomic lines, the molecular isotopologues (e.g., $^{13}$CH, H$_2^{18}$O, H$_2^{17}$O, see \citealt{mul23} for their detection) are well separated in frequency. This removes the problem of possible variance of the elemental isotopic ratios from their terrestrial values, encountered in the fitting of optical absorbers when isotopes are blended (e.g., \citealt{web25}).

Lastly, we have considered above that the relative strengths of the components of the CH hyperfine structure follow their statistical weights, given in Table~\ref{tab:spectro}. To test this assumption, we have fitted the spectra again with the same procedure as explained in Sec.\,\ref{sec:gaussfit}, but without the bulk velocity offset between CH and H$_2$O and with the relative strengths of the hyperfine components of CH set as free parameters. Given the limits of the signal-to-noise ratio of the spectra and the number of velocity components, this fitting exercise could only be performed against PKS\,1830$-$211(SW) in order to avoid large degeneracies between the placement and amplitude of the velocity components and relative strengths of the hyperfine components. Again, we used the spectra observed on May 5, 2014 together with H$_2^{18}$O to guide the fit around the saturated part of the H$_2$O absorption. We have fixed the strength of the strongest hyperfine component of the 533~GHz transitions to 0.83, and found the following results for the others, with the expected statistical weights given in parentheses: $0.37 \pm 0.02$ (vs 0.33) and $0.15 \pm 0.04$ (vs 0.17) for the 533~GHz transitions, and $0.86 \pm 0.04$ (vs 0.83), $0.37 \pm0.02$ (vs 0.33), and $0.15 \pm 0.01$ (vs 0.17) for the 537~GHz transitions, in order of decreasing strengths, respectively \footnote{The fit also provided $f_c = 0.946 \pm 0.002$, and $\gamma_\tau = 8.6 \pm 0.4$, with $\chi^2  = 1.4$}. Within uncertainties, these results are consistent with the values of the statistical weights, and therefore, the initial assumption is validated, at least for this data set. In turn, this is also a good indication that the fitted velocity structure is not very different from the true one, and whatever deviations exist, it is not causing large discrepancies in the derived statistical weights.

More generally, we have explored with RADEX \citep{vdtak07} the excitation of submillimeter lines of CH under standard interstellar conditions, using hyperfine-resolved collision data from \cite{dag18}. In contrast to the radio wavelength transitions at $\sim 3$~GHz which are known for their anomalous excitation (see, e.g., \citealt{jac24} and references therein), we find that the excitation temperatures of the submillimeter transitions remain well coupled to the temperature of the cosmic microwave background (e.g., within a few tenths of K from $T_\mathrm{CMB}$ for a kinetic temperature of 50~K and an H$_2$ volume density up to $10^4$~cm$^{-3}$).

\subsubsection{Properties of the absorbers}

The physical and chemical conditions of the absorbing gas in the PKS\,1830$-$211(SW) line of sight have been investigated in detail (e.g., \citealt{hen09, mul11,mul13, mul14a,mul16,mul17, sch15}) and are found to be comparable to those of translucent clouds in the Milky Way. For the PKS\,1830$-$211(NE) line of sight, the conditions are not as well known, mainly because the molecular absorption there is weaker -- column densities are typically one order of magnitude lower than those along PKS\,1830$-$211(SW). However, the properties of the NE absorption are clearly reminiscent of a diffuse component, as evidenced by 1) the prominence of the H\,I absorption spectrum in the NE line of sight (\citealt{koo05,com21}), 2) the enhancement of molecular tracers of low H$_2$-fraction gas, such as ArH$^+$ \citep{mul15}, OH$^+$ and H$_2$O$^+$ \citep{mul16}, H$_2$Cl$^+$ \citep{mul14b}, and CH$^+$ \citep{mul17}, and 3) the conditions traced by hydrogen radio recombination lines \citep{emi23}. The conditions in the absorber toward B\,0218$+$357(SW) are believed to be intermediate between those in the NE and SW lines of sight toward PKS\,1830$-$211, dominated by a diffuse medium \citep{hen05,wal16,wal19}.

For a given telescope, the sensitivity of absorption spectroscopy depends on the brightness of the background continuum illumination and the opacity of the foreground material. Both the absorption systems PKS\,1830$-$211 and B\,0218$+$357 present very favorable properties. The quasars are bright at submillimeter wavelengths ($\sim 0.1 - 1$~Jy) and the absorbers have relatively high column densities of absorbing gas (N(H$_2$) $\sim 10^{21-22}$~cm$^{-2}$). In fact, the column density of water in PKS\,1830$-$211(SW) is so large that the 557-GHz line becomes highly saturated, forcing us to use the H$_2^{18}$O isotopologue to sample the saturated region. On the other hand, the PKS\,1830$-$211(NE) and B\,0218$+$357(SW) sightlines are not saturated, but they still present relatively strong lines. In particular, the PKS\,1830$-$211(NE) sightline is very favorable for three reasons. First, its absorption profile is made up of multiple ($\gtrsim 10$), narrow (a few km\,s$^{-1}$ wide), and relatively well separated velocity components, which is highly favorable for velocity offset measurements. For B\,0218$+$357(SW), the absorption profile is more compact, spread over a smaller velocity range, and composed of only a handful of velocity components, partially overlapping. The number of velocity components in these lines of sight is thus sufficiently large to put strong constraints, but still not too large to severely affect the fitting procedure with fit ambiguities and convergence issues (see, e.g., \citealt{mur22}). Secondly, the width of the absorption components is also very favorable for obtaining a high precision ($<$ km\,s$^{-1}$) when comparing line kinematics. Third, the time variability of the absorption profile toward PKS\,1830$-$211(NE) allows us to multiply the number of independent measurements with multi-epoch observations. This improves statistics and reduces potential systematic effects when fitting CH and H$_2$O with the same profile.

\subsubsection{Observational setup}

The heterodyne technique provides high frequency resolution ($\delta \nu / \nu \sim 10^{-6}$ in our case) and the instrumental radio frequency bandpass is calibrated with high precision. Since the respective absorption features along the NE and SW lines of sight to PKS\,1830$-$211 do not overlap in velocity, we can also compare both spectra and check that there is no contamination by atmospheric lines or any bandpass artifacts. 

The CH and H$_2$O lines are observed simultaneously within the same frequency tuning, with the same atmospheric conditions, the same correlator setup, and the same velocity resolution. Data calibration was performed in parallel with the same procedure. Any instrumental effects are therefore minimized.

Most importantly, the observation of the CH and H$_2$O lines, simultaneous and close in frequency within the same tuning, removes the two most severe sources of known systematics toward PKS\,1830$-$211 (which may also apply to some extent to B\,0218+357): the time variability of the absorption profile, which hampers the comparison of spectra taken at different epochs \citep{mul08,sch15,mul23}, and the frequency-dependent morphology of the background quasar, which can affect the comparison of absorption lines taken at different frequencies (e.g., \citealt{mar13, kan15}).

\subsubsection{Co-spatiality of CH and H$_2$O}

Herschel observations of the same submillimeter lines of CH and H$_2$O in absorption in diffuse sightlines of the Milky Way find that both species are closely correlated and arise from the same gas component (\citealt{qin10,ger10,fla13}). The lines connect to the ground-state energy level and have similar subthermal excitation for both species, mostly coupled with photons from the cosmic microwave background ($T_\mathrm{CMB}=4.60$~K at $z=0.68$ and $T_\mathrm{CMB} = 5.14$~K at $z=0.89$, respectively) at relatively low H$_2$ volume density of a few $10^3$~cm$^{-3}$ or less.

The excellent correlation between CH and H$_2$O opacities was highlighted by \cite{mul23} in both PKS\,1830$-$211 SW and NE sightlines. Although we cannot formally prove that line kinematics do not suffer from chemical differentiation within individual velocity components and that the two species are co-spatial, the overall excellent correlation between CH and H$_2$O, holding in both absorbers and even in spite of time variations of the absorption profile resulting from morphological changes in the background quasar, gives strong evidence that both species are tracing the same gas component.

\section{Conclusions} \label{Conclusions}

We have used ALMA observations of CH and H$_2$O submillimeter lines in the molecular absorbers toward the quasars PKS\,1830$-$211 ($z_{\rm abs}=0.89$) and B\,0218$+$357 ($z_{\rm abs}=0.68$), respectively, to test the invariance of the fine structure constant, $\alpha$, and the proton-to-electron mass ratio, $\mu$, at look-back times of nearly half the present age of the Universe with respect to their values, here and today on Earth. Our observational setup minimizes known specific systematics of radio absorbers, such as the time variations of the absorption line profile and the frequency-dependent morphology of the background quasars, as well as instrumental and calibration errors. Based on the average bulk velocity offsets between the CH and H$_2$O absorption profiles and the independent upper limits on $\Delta \mu / \mu$ previously obtained in these systems, we derive $|\Delta \alpha / \alpha| < 0.55$~ppm and 1.5~ppm in the direction of PKS\,1830$-$211 and B\,0218$+$357, respectively, at $3\sigma$ confidence level. To the best of our knowledge, these are the most stringent constraints on cosmological variations of $\alpha$, so far.

This work highlights the strength of distant radio molecular absorbers for testing the invariance of fundamental constants and, at the same time, the uniqueness of PKS\,1830$-$211 and B\,0218$+$357, as we are not aware of any other sources capable of providing such deep constraints using a similar setup.

\begin{acknowledgement}
We thank the referee for her/his contructive comments.
This paper makes use of data from the following ALMA projects: \\
ADS/JAO.ALMA\#2011.0.00405.S (data from April 2012), \\
ADS/JAO.ALMA\#2012.1.00056.S (data from May 2014), \\
ADS/JAO.ALMA\#2013.1.00020.S (data from July 2014), \\
ADS/JAO.ALMA\#2015.1.00075.S (2016 data) for PKS\,1830$-$211, \\
and ADS/JAO.ALMA\#2016.1.00031.S for B\,0218$+$357. 
ALMA is a partnership of ESO (representing its member states), NSF (USA) and NINS (Japan), together with NRC (Canada) and NSC and ASIAA (Taiwan), in cooperation with the Republic of Chile. The Joint ALMA Observatory is operated by ESO, AUI/NRAO and NAOJ. This research has made intensive use of the NASA's Astrophysics Data System.
\end{acknowledgement}


\begin{thebibliography}{}

  
\bibitem[Altman et al.(2025)]{alt25}{Altman, A., T\'obi\'as, R., Bogomolov, A.S., et al., 2025, \apjs, 280, 47}
\bibitem[Bagdonaite et al.(2013a)]{bag13a}{Bagdonaite, J., Jansen, P., Henkel, C., et al., 2013a, Science, 339, 46}
\bibitem[Bagdonaite et al.(2013b)]{bag13b}{Bagdonaite, J., Dapr\`a, M., Jansen, P., et al., 2013b, \prl, 111, 231101}
\bibitem[Bahcall \& Schmidt(1967)]{bah67}{Bahcall, J. N. \& Schmidt, M., 1967, \prl, 19, 1294}
\bibitem[Brown \& Evenson(1983)]{bro83}{Brown, J. M. \& Evenson, K. M., 1983, \apj, 268, 51}
\bibitem[Calmet \& Fritzsch(2006)]{cal06}{Calmet X. \& Fritzsch H., 2006, Europhys. Lett., 76, 1064}
\bibitem[Cazzoli et al.(2009)]{caz09}{Cazzoli, G., Puzzarini, C., Harding, M.\,E., \& Gauss, J., 2009, Chem. Phys. Lett., 473, 21}
\bibitem[Combes et al.(2021)]{com21}{Combes, F., Gupta, N.,  Muller, S., et al., 2021, \aap, 648, 116}
\bibitem[Dagdigian(2018)]{dag18}{Dagdigian, P. J., 2018, \mnras, 475, 5480}
\bibitem[de Nijs et al.(2012)]{nij12}{de Nijs, A. J., Ubachs, W. \& Bethlem, H. L., 2012, \pra, 86, 2501}
\bibitem[Dirac(1937)]{dir37}{Dirac, P. A. M, 1937, \nat, 139, 323}
\bibitem[Drinkwater et al.(1998)]{dri98}{Drinkwater, M. J., Webb, J. K., Barrow, J. D., \& Flambaum, V. V., 1998, \mnras, 295, 457}
\bibitem[Emig et al.(2023)]{emi23}{Emig, K. L., Gupta, N., Salas, P., et al., 2023, \apj, 994, 93}
\bibitem[Evans et al.(2014)]{eva14}{Evans, T. M., Murphy, M. T., Whitmore, J. B., et al., 2014, \mnras, 445, 128}
\bibitem[Flagey et al.(2013)]{fla13}{Flagey, N., Goldsmith, P. F., Lis, D. C., et al., 2013, \apj, 762, 11}
\bibitem[Frieman et al.(2008)]{fri08}{Frieman, J. A., Turner, M. S., \& Huterer, D., 2008, \araa, 46, 385}
\bibitem[Frye et al.(1997)]{fry97}{Frye, B., Welch, W. J., \& Broadhurst, T., 1997, \apj, 478, 25}
\bibitem[G\'erin et al.(2010)]{ger10}{G\'erin, M., de Luca, M., Goicoechea, J. R., et al., 2010, \aap, 521, 16}
\bibitem[G\'erin et al.(2016)]{ger16}{G\'erin, M., Neufeld, D., \& Goicoechea, J., 2016, ARA\&A, 54, 181}
\bibitem[Golubiatnikov et al.(2006)]{gol06}{Golubiatnikov, G.\,Y., Markov, V.\,N., Guarnieri, A., \& Kn{\"o}chel, R., 2006, J. Mol. Spectrosc., 240, 251}
\bibitem[Hart \& Chluba(2020)]{har20}{Hart, L. \& Chluba, J., 2020, \mnras, 493, 3255}
\bibitem[Henkel et al.(2005)]{hen05}{Henkel, C., Jethava, N., Kraus, A., et al., 2005, \aap, 440, 893}
\bibitem[Henkel et al.(2009)]{hen09}{Henkel, C., Menten, K. M., Murphy, M. T., et al., 2009, \aap, 500, 725}
\bibitem[H\"ogbom(1974)]{hog74}{H\"ogbom, J. A., 1974, \aaps, 15, 417}
\bibitem[Jacob et al.(2024)]{jac24}{Jacob, A. M., Nandakumar, M., Roy, N., et al., 2024,  \aap, 692, 164}
\bibitem[Kanekar(2011)]{kan11}{Kanekar, N., 2011, \apj, 728, 12}
\bibitem[Kanekar et al.(2015)]{kan15}{Kanekar, N., Ubachs, W., Menten, K. M., et al., 2015, \mnras, 448, 104}
\bibitem[King et al.(2012)]{kin12}{King, J. A., Webb, J. K., Murphy, Michael T., et al., 2012, \mnras, 422, 3370}
\bibitem[Koopmans \& de Bruyn(2005)]{koo05}{Koopmans, L. V. E. \& de Bruyn, A. G., 2005, \mnras, 360, L6}
\bibitem[Lee et al.(2023)]{lee23}{Lee, N., Ali-Ha\"imoud, Y., Sch\"oneberg, N., \& Poulin, V., 2023, \prl, 130, 161003}
\bibitem[Mahieu et al.(2012)]{mah12}{Mahieu, S., Maier, D., Lazareff, B., et al., 2012, IEEE Trans. Terahertz Sci. Technol., 2, 29}
\bibitem[Martins(2017)]{mar17}{Martins, C. J. A. P., 2017, Reports on Progress in Physics, 80, 126902}
\bibitem[Mart\'i-Vidal et al.(2013)]{mar13}{Mart\'i-Vidal, I., Muller, S., Combes, F., et al., 2013, \aap, 558, 123}
\bibitem[Mart\'i-Vidal et al.(2014)]{mar14}{Mart\'i-Vidal, I., Vlemmings, W., Muller, S., \& Casey S., 2014, \aap, 563, 136}
\bibitem[Menten \& Reid(1996)]{men96}{Menten, K. M. \& Reid, M. J., 1996, \apj, 465, 99}
\bibitem[Mohr et al.(2025)]{moh25}{Mohr, P. J., Newell, D. B., Taylor, B. N., \& Tiesinga, E., 2025, Rev. Mod. Phys. 97, 025002}
\bibitem[Molaro et al.(2013)]{mol13}{Molaro, P., Centuri\'on, M., Whitmore, J. B., et al., 2013, \aap, 555, 68}
\bibitem[Muller et al.(2006)]{mul06}{Muller, S., Gu\'elin, M., Dumke, M., et al., 2006, \aap, 458, 417}
\bibitem[Muller et al.(2007)]{mul07}{Muller, S., Gu\'elin, M, Combes, F., \& Wiklind, T., 2007, \aap, 468, 53}
\bibitem[Muller \& Gu\'elin(2008)]{mul08}{Muller, S. \& Gu\'elin, M., 2008, \aap, 491, 739}
\bibitem[Muller et al.(2011)]{mul11}{Muller, S., Beelen, A., Gu\'elin, M., et al., 2011, \aap, 535, 103}
\bibitem[Muller et al.(2013)]{mul13}{Muller, S., Beelen, Black J. H., et al., 2013, \aap, 551, 109}
\bibitem[Muller et al.(2014a)]{mul14a}{Muller, S., Combes, F., Gu\'elin, M., et al., 2014a, \aap, 566, 112}
\bibitem[Muller et al.(2014b)]{mul14b}{Muller, S., Black, J. H., Gu\'elin, M., et al., 2014b, \aap, 566, 6}
\bibitem[Muller et al.(2016)]{mul16}{Muller, S., M\"uller, H. S. P., Black, J. H., et al., 2016, \aap, 595, 128}
\bibitem[Muller et al.(2017)]{mul17}{Muller, S.,  M\"uller, H. S. P., Black, J. H., et al., 2017, \aap, 606, 109}
\bibitem[Muller et al.(2020)]{mul20}{Muller, S., Jaswanth, S., Horellou, C., \& Mart\i-Vidal, I., 2020, \aap, 641, 2}
\bibitem[Muller et al.(2021)]{mul21}{Muller, S., Ubachs, W., Menten, K. M., Henkel, C., \& Kanekar, N., 2021, \aap, 652, 5}
\bibitem[Muller et al.(2023)]{mul23}{Muller, S., Mart\'i-Vidal, I., Combes, F., et al., 2023, \aap, 674, 101} 
\bibitem[M\"uller et al.(2015)]{mul15}{M\"uller, H. S. P., Muller, S., Schilke, P., et al., 2015, \aap, 582, 4}
\bibitem[Murphy et al.(2001)]{mur01}{Murphy, M. T., Webb, J. K., Flambaum, V. V., et al., 2001, \mnras, 327, 1208}
\bibitem[Murphy et al.(2003)]{mur03}{Murphy, M. T., Webb, J. K., \& Flambaum, V. V., 2003, \mnras, 345, 609}
\bibitem[Murphy et al.(2008)]{mur08}{Murphy, M. T., Flambaum, V. V., Muller, S. \& Henkel, C., 2008, Science, 320, 1611}
\bibitem[Murphy \& Cooksey(2017)]{mur17}{Murphy, M. T. \& Cooksey, K. L., 2017, \mnras, 471, 4930}
\bibitem[Murphy et al.(2022)]{mur22}{Murphy, M. T., Molaro, P., Leite, A. C. O., 2022, \aap, 658, 123}
\bibitem[O'Dea et al.(1992)]{ode92}{O’Dea, C. P., Baum, S. A., Stanghellini, C., et al., 1992, \aj, 104, 1320}
\bibitem[Patnaik et al.(1993)]{pat93}{Patnaik, A. R., Browne, I. W. A., King, L. J., et al., 1993, MNRAS, 261, 435}
\bibitem[Qin et al.(2010)]{qin10}{Qin, S.-L., Schilke, P., Comito, C., et al., 2010, \aap, 521, 14}
\bibitem[Rahmani et al.(2013)]{rah13}{Rahmani, H., Wendt, M., Srianand, R., et al., 2013, \mnras, 435, 861}
\bibitem[Riess et al.(1998)]{rie98}{Riess, A. G., Filippenko, A. V., Challis, P., et al., 1998, \aj, 116, 1009}
\bibitem[Rydbeck et al.(1973)]{ryd73}{Rydbeck, O. E. H., Elld\'er, J., \& Irvine, W. M., 1973, \nat, 246, 466}
\bibitem[Savedoff(1956)]{sav56}{Savedoff, M. P., 1956, \nat, 178, 688}
\bibitem[Sch\"oneberg \& Vacher(2025)]{sch25}{Sch\"oneberg, N. \& Vacher, L., 2025, Journal of Cosmology and Astroparticle Physics, 03, 004}
\bibitem[Schulz et al.(2015)]{sch15}{Schulz, A., Henkel, C., Menten, K. M., et al., 2015, \aap, 574, 108}
\bibitem[Sekiguchi \& Takahashi(2021)]{sek21}{Sekiguchi, T. \& Takahashi, T., 2021, Phys. Rev. D, 103, 083507}
\bibitem[Sheffer et al.(2008)]{she08}{Sheffer, Y., Rogers, M., Federman, S. R., et al., 2008, \apj, 687, 1075}
\bibitem[Subrahmanyan et al.(1990)]{sub90}{Subrahmanyan, R., Narasimha, D., Pramesh Rao, A., \& Swarup, G., 1990, \mnras, 246, 263}
\bibitem[Swings \& Rosenfeld(1937)]{swi37}{Swings, P. \& Rosenfeld, L., 1937, \apj, 86, 483}
\bibitem[Truppe et al.(2013)]{tru13}{Truppe, S., Hendricks, R. J., Tokunaga, S. K., et al., 2013, Nature Communications, 4, 2600}
\bibitem[Truppe et al.(2014)]{tru14}{Truppe, S., Hendricks, R. J., Hinds, E. A., \& Tarbutt, M. R., 2014, \apj, 780, 71}
\bibitem[Turner \& Zuckerman(1974)]{tur74}{Turner, B. E. \& Zuckerman, B., 1974, \apj, 187, 59}
\bibitem[Uzan(2025)]{uza25}{Uzan J.-P., 2025, Living Reviews in Relativity, 28, 6}
\bibitem[van der Tak et al.(2007)]{vdtak07}{van der Tak, F. F. S., Black, J. H., Sch\"oier, F. L., Jansen, D. J., van Dishoeck, E. F., 2007, \aap, 468, 627}
\bibitem[Wallstr\"om et al.(2016)]{wal16}{Wallstr\"om, S. H. J., Muller, S., \& Gu\'elin, M., 2016, \aap, 595, 96}
\bibitem[Wallstr\"om et al.(2019)]{wal19}{Wallstr\"om, S. H. J., Muller, S., Roueff, E., et al., 2019, \aap, 629, 128}
\bibitem[Webb et al.(1999)]{web99}{Webb, J. K., Flambaum, V. V., Churchill, C. W., et al., 1999, \prl, 82, 884}
\bibitem[Webb et al.(2001)]{web01}{Webb, J. K., Murphy, M. T., Flambaum, V. V., et al., 2001, \prl, 87, 091301}
\bibitem[Webb et al.(2011)]{web11}{Webb, J. K., King, J. A., Murphy, M. T., et al., 2011, \prl, 107, 1101}
\bibitem[Webb et al.(2025)]{web25}{Webb, J. K., Lee, C.-C., Milakovi\'c, D., et al., 2025, \mnras, 539, 1}
\bibitem[Whitmore \& Murphy(2015)]{whi15}{Whitmore, J. B. \& Murphy, M. T., 2015, \mnras, 447, 446}
\bibitem[Wiklind \& Combes(1995)]{wik95}{Wiklind, T. \& Combes F. 1995, \aap, 299, 382}
\bibitem[Wiklind \& Combes(1996)]{wik96}{Wiklind T. \& Combes F. 1996, \nat, 379, 139}
\bibitem[Wiklind \& Combes(1998)]{wik98}{Wiklind T. \& Combes F. 1998, \apj, 500, 129}
\bibitem[Wilczynska et al.(2015)]{wil15}{Wilczynska, M. R., Webb, J. K., King, J. A., 2015, \mnras, 454, 3082}
\bibitem[Wilczynska et al.(2020)]{wil20}{Wilczynska, M. R., Webb, J. K., Bainbridge, M., et al., 2020, Science Advances, 6, 9672}

\end{thebibliography}
\end{document}